\documentclass[runninghead]{llncs}

\usepackage[ruled,linesnumbered]{algorithm2e}

\usepackage{wrapfig}
\def\BibTeX{{\rm B\kern-.05em{\sc i\kern-.025em b}\kern-.08em
    T\kern-.1667em\lower.7ex\hbox{E}\kern-.125emX}}
    
\usepackage{CJKutf8}  

\usepackage{xcolor}
\usepackage{pgfplots}
\usepackage{tikz}
\usepackage{comment}

\usepackage{filecontents}

\usepackage{pifont} %used for \ding

\usepackage{fancyhdr} %add page number
\usepackage{hyperref}
\usepackage{url}

\usepackage{fbox} %use box
\usepackage{tikz} % to draw some self-contained figs
\usepackage[most]{tcolorbox}

\usepackage{multirow}
\usepackage{cite}
\usepackage{amsmath,amssymb,amsfonts}

\usepackage{textcomp}
\usepackage{xcolor}

\usepackage{subfigure}
\usepackage{pgfplots}
\usepackage{filecontents}

\usepackage{graphicx}
\usepackage{graphicx,subfigure}

\usepackage{float}

\definecolor{bblue}{HTML}{4F81BD}
\definecolor{rred}{HTML}{C0504D}
\definecolor{ggreen}{HTML}{9BBB59}
\definecolor{ppurple}{HTML}{9F4C7C}

\usepackage{listings}

\usepackage{blindtext}
\usepackage{etoolbox,xstring,mfirstuc,textcase}
\usepackage{booktabs}
\usepackage{pgfplots}

\usepackage{tabularx,booktabs,makecell,multirow, caption}
\usepackage{enumitem} % to control Itemization spacing

\newcolumntype{L}{>{\arraybackslash}X}

\usepackage{xcolor}
\usepackage{tikz}
\usepackage{pgfplots}

\definecolor{findOptimalPartition}{HTML}{D7191C}
\definecolor{storeClusterComponent}{HTML}{FDAE61}
\definecolor{dbscan}{HTML}{ABDDA4}
\definecolor{constructCluster}{HTML}{2B83BA}

\usepackage[utf8]{inputenc}
\usepackage{listings}
\begin{document}
%=================================================   
\title{FTX Collapse: A Ponzi Story}
%=================================================   
%=================================================   

\author{Shange Fu\inst{1}, Qin Wang\inst{2}, Jiangshan Yu\inst{1}, Shiping Chen\inst{2}}
\authorrunning{S. Fu, Q. W, J Yu, S. Chen} % Part of LEFT running header
\titlerunning{Rational Ponzi Games}% Part of RIGHT running header

\institute{
Monash University, Australia
\\
\and
CSIRO Data61, Australia
}

%======================================
\maketitle
%====================================== 

\begin{abstract}
FTX used to be the third-largest centralised exchange (CEX) in crypto markets, managing over \$10B in daily trading volume before its downfall. Such a giant, however, failed to avoid the fate of mania, panic, and crash. In this work, we revisit the FTX's crash by telling it as a Ponzi story. In regards to why FTX could not sustain this Ponzi game, we extract and demonstrate the three facilitators of the FTX collapse, namely, \textit{FTT}, \textit{leverage}, and \textit{diversion}. The unfunctionality in each factor can iteratively magnify the impact of damages when the panic is triggered.
Rooted in the unstable ground, FTX eventually suffered insolvency and rapidly crashed in Nov. 2022.
The crisis of FTX is not an isolated event; it consequently results in the collapse of a chain of associated companies in the entire crypto market.
Recall this painful experience, we discuss possible paths for a way forward for both CeFi and DeFi services.


\end{abstract}

%=================================================   
\section{Introduction}
%=================================================  
FTX, before collapsing, was the third-largest cryptocurrency exchange that handles over \$10 active trading volume~\cite{cmc}. FTX was created by its founder Sam Bankman-Fried (SBF in short) in 2019 and experienced a skyrocketing development within three years. Besides this influential trading platform, SBF also created a hedge fund called Alameda Research that manages \$14.6B assets (\$6B from FTT\footnote{FTT is the native token of the crypto derivatives created by FTX that is launched on May 8, 2019. At the time of writing, FTX token price has dropped 62X from \$79.53 (9/10/2021) to \$1.29 (11/29/2022)~\cite{cmc}.} and other collateral while \$8B liabilities \cite{CoinDesk1}). However, FTX struck a deal for Binance and halted all non-fiat customer withdrawals. Later, FTX
declared bankruptcy on Nov. 11, 2022, which marks the collapse of the entire empire. The negative impact rapidly spreads over entire financial markets. Most mainstream cryptocurrency prices have declined more than 30\% within hours. A large number of crypto-projects suffer inevitable catastrophes due to their close connection with FTX (e.g., Solana \cite{solanaftx}\cite{ftxbacked}). Many traditional finance agencies that have relevant trades, such as hedge funds, also confront a huge amount of monetary loss. U.S. Department of Justice and federal agencies thereby started to investigate the ins and outs of the crash \cite{usdepartment}. %contagiously

Back to FTX's origin, we can observe a common pattern used in crypto projects: issue \textit{non-collateral tokens} (primarily in the forms of ERC-20 tokens \cite{erc20} in Ethereum ecosystem \cite{wood2014ethereum}) without any backed-up assets. This phenomenon is also known as ICO (short for Initial Coin Offering) which was prevalent since 2017. Not surprisingly, FTX issued its native token FTT in 2019, raising a significant of high-liquidity crypto assets including BTC and ETH. However, such a design pattern implicitly inherits the Ponzi nature \cite{basu2014ponzi}, in which the asset manager always borrows new money to pay off the old debts until the game collapses. Typically, the issuance of a new token is costless (equiv. without collateral).

The reasons behind the FTX collapse are not pioneering, though 
indeed it was facilitated in several innovative ways by DeFi protocols.
In fact, apart from pure scams in crypto markets, many means from traditional financial markets have also been applied in today's DeFi games as an add-on, with \textit{leverage} being the most popular one.
Rather than its classic connotation on financial derivatives like futures and options, leverage refers to the iterative magnification of the impacts of one type of created tokens (FTT in this paper).
One FTT token can yield, perhaps, 100X or much more value after traveling across several centralised exchanges and DeFi protocols \cite{werner2021sok}. Another important factor is \textit{divert}, referring to the misappropriation of reserves. In particular, current crypto-markets are absent of formal regulation, which has spawned many types of CeFi/DeFi services without rigorous background checks or authentication.
There is an unbelievable fact that crypto-users, unfortunately, are keen on this kind of game due to its potential for extremely high-return.
A similar story happens on FTX, and with a high probability, will happen repeatedly in the future. This motivates us to explore a series of decisive factors that may alter the ending of stories.

\smallskip
\noindent\textbf{Contribution.} We explore the reasons that cause FTX's collapse by retelling the story in the context of the Ponzi model. We first present the fundamentals of Ponzi and, more importantly, a rational Ponzi game as the baseline of our explanation (\textcolor{magenta}{Sec.\ref{sec:ponzi}}). Based on this theoretical model, we decompose three major phases of FTX's growth and accordingly analyze the root reasons that break the game rules (\textcolor{magenta}{Sec.\ref{sec:ftx}}). We show that the failure of FTX is not an accident but, instead, caused by its severe violations against a rational model.
Further, we provide discussion and advice
on potential aspects, namely regulation and transparency, for future endeavors (\textcolor{magenta}{Sec.\ref{sec:discuss}}).
% Hopefully, this work may push users to rethink the ways of designing much more stable protocols in the crypto space (e.g., \cite{jumpto}\cite{stablecoins}).

\smallskip
\noindent\textbf{Related Work.} Nansen \cite{nansen} provides a research analysis that captures on-chain data from May 2019 - Sep 2022, focusing on both FTT and Alameda activities. Ramirez \cite{daliaftx} sorts out the timeline and milestones of the FTX crash. Jakub \cite{pathforwd} aims to point out the future paths for centralised exchanges.  Besides, many influential companies (e.g., CoinDesk \cite{CoinDesk}, Investopedia \cite{investopedia}) and other mainstream media (e.g., Forbes \cite{forbes}, Wiki \cite{wiki}), also put their focus on such a historical event by recalling, analysing, and summarising its ins and outs.
However, most of the cryto-collapse research, reports, and posts \cite{jumpto}\cite{stablecoins}
(topics also covering LUNA-UST's depeg)
% ,Three Arrows Capital's failure, etc.)
are based on facts and explicit data without the in-depth abstraction of easy-understanding models or insights, which is the key matter that we aim to deliver in this paper.  

%=================================================
\section{What is Ponzi?}
\label{sec:ponzi}
%================================================

The Ponzi game, commonly referred to as the Ponzi scheme, is named after Charles Ponzi, who duped investors in the 1920s with a postage stamp speculation scheme.
Ponzi is essentially a financial protocol that old liability should be financed by issuing new debt.
For those who are deliberately deceptive, Ponzi scheme organizers often promise to invest your money and generate high returns with little or no risk.
But in practical Ponzi schemes, fraudsters do not invest money.
Instead, they use it to pay those who invested earlier and may keep some for themselves.
Ponzi schemes normally share common characteristics, and U.S. Securities and Exchange Commission~\cite{ponzi_gov} raises ``red flags'' for these warning signs:
(i) high returns with little or no risk;
(ii) overly consistent returns;
(iii) unregistered investments;
(iv) unlicensed sellers;
(v) secretive, complex strategies;
(vi) issues with paperwork; and
(vii) difficulty receiving payments.

\begin{figure}[H]
	\centering
    \includegraphics[scale=0.665]{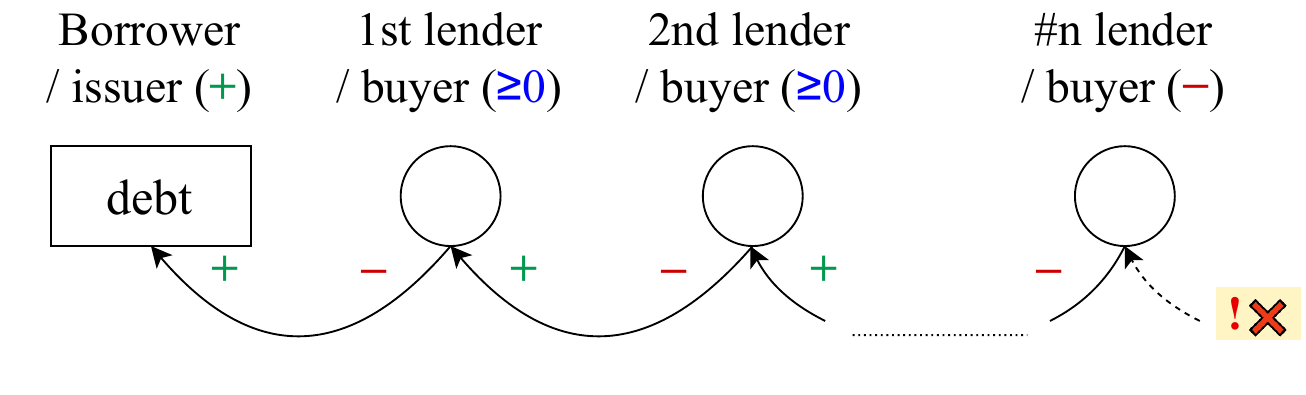}
	\vspace{-0.5cm}
	\caption{Example: How a Ponzi scheme works.}
	\label{fig:ponzi}
 	\vspace{-0.6cm} 
\end{figure}

We then demonstrate how a Ponzi scheme works (and how it may be broken) in general in Fig.\ref{fig:ponzi}.
The green plus sign and the red minus sign represent cash inflows and cash outflows, respectively.
The values in parentheses following each participant represent their corresponding utility.
In a Ponzi scheme, the borrower or the issuer can always benefit from the game, and participants can reach a \textit{no-less-than-zero} utility before the breaking of Ponzi, only the last wave of escapees will suffer losses.
Therefore, a broken Ponzi scheme is a \textbf{\textit{zero-sum}} game.
%Edit the figure here.
% https://drive.google.com/file/d/1P-sIFJB8re-tA93vH0k21mq5bQZpHSEZ/view?usp=sharing

\smallskip
\noindent\textbf{Rational Ponzi Game}.
Running a Ponzi, however, does not necessarily imply that any participant is in any sense losing out, as long as the game can be perpetually rolled over. Economists call such realization as a \textit{rational Ponzi game}~\cite{ponzi_rational}. A rational Ponzi game refers to a type of sustainable game that satisfies two major factors of \textit{permanently increasing} and \textit{no harm to any party}. It requires monetary investment to the game should continuously grow without degrading the situation of participants. Economically, the game is measured by a utility function (a.k.a. pay-off function) based on the net cash inflows of lenders $ \mathcal{I}_s$, the borrower's net indebtedness $ \mathcal{D}_{T}$, the discount factor $\Gamma(T)$ (calculated based on the interest rate $r_t$), as well as the time-based counters $T$ and $s$. The key principle is summaries as \textit{all future assets in the form of present values can always be redeemed deposited assets values.} It can be summarised in a loose mathematical model \cite{fu2022rational} which is defined as:

\begin{equation}
\begin{aligned}
\Gamma(T)  \mathcal{D}_{T} =& \sum_{s=1}^{T} \Gamma(s) \mathcal{I}_{s}, \\
\textit{where}\,\,  \Gamma(s) \equiv & \prod_{j=1}^{s}\left(1+r_{j}\right)^{-1}.
\end{aligned} 
\end{equation}

%\begin{equation*}
% \begin{aligned}
%    | Pr[\mathcal{A}(c)=b: &CM1\gets RVer, CM2\gets EVer:(m_0,m_1)\gets \mathcal{A}(CM1,CM2); \\
%     & b\gets \{0,1\};c\gets HDec(CM1,CM2,m_b) ] -\frac{1}{2}| \leq negl(\lambda)\\
% \end{aligned} 
%\end{equation*}

The model is simulated as a sequential of debt transactions \cite{blanchard1992dynamic} that need to hold two conditions that are separately fit for different participants: firstly (\textcolor{magenta}{$\natural$-\ding{172}}), the net present indebtedness value (\textit{left}) should be always positive for borrowers; secondly (\textcolor{magenta}{$\natural$-\ding{173}}), the game should not hurt any participating parties in the game (a.k.a. Pareto improvement \cite{pareto}). These two properties guarantee a perpetually operating economical game in the context of permissionless environments. This also explains the name of \textit{rational Ponzi} in which all the rational players will, at least, lose nothing in the game.

%=================================================
\section{The FTX Collapse}
\label{sec:ftx}
%================================================

The story of FTX starts with its founder SBF. Though there already existed hundreds of CEXes and strong competitors such as Coinbase or Binance, SBF still decided to launch a new crypto exchange FTX in April 2019.
The growth of FTX was sped up by cooperating with a good market maker Alameda to provide liquidity.
Interestingly, Alameda is also founded by SBF.
The great collaboration between FTX and Alameda made FTX rapidly grow up to the global Top3 CEX, and made the valuation of FTX balloon in size to \$32 billion as of January 2022~\cite{pitchbook}.
With such success, FTX naturally launched its self-issued token FTT just like most centralised exchanges. However, with all the familiar stages in order - token boom and a fad for it, then leverage and diversion, then problem revealed and panic followed close behind, FTX finally fell into its crash.
In this section, we decompose this disastrous speculation and demonstrate three facilitators that explain the FTX collapse, namely \textit{FTT}, \textit{leverage}, and \textit{diversion}.

% You can edit this figure here.
%https://drive.google.com/file/d/1a3urDMwdQco3fG6Ikq7RHl-Pci9m8cF5/view?usp=sharing
\begin{figure*}[!hbt]
	\centering
    \includegraphics[width=\linewidth]{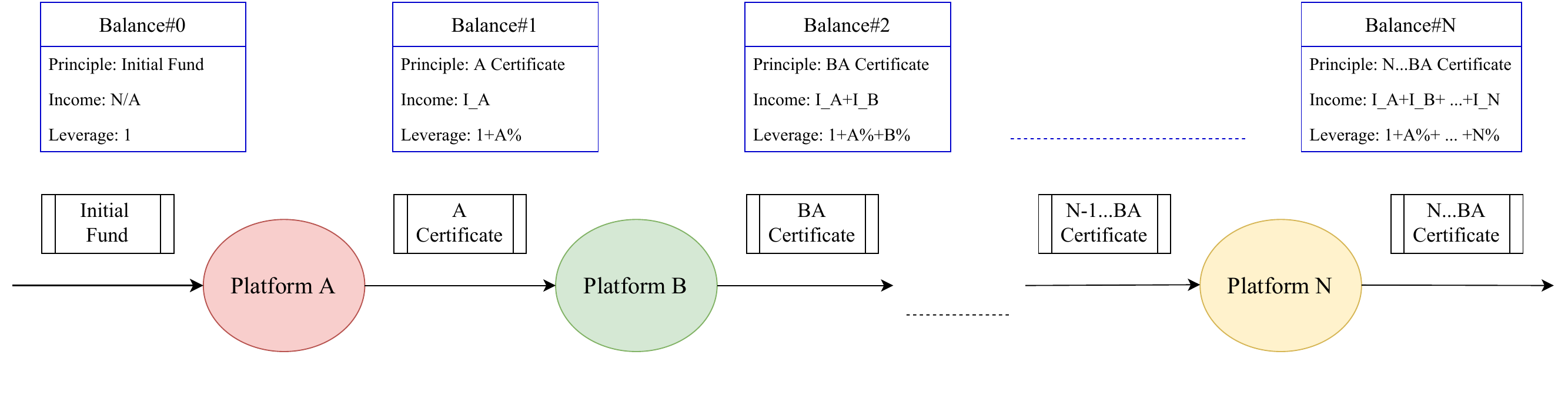}
	\caption{The Leverage Mechanisms}
	\label{fig:leverage}
\end{figure*}

\smallskip
\noindent\textbf{FTT.}
\textit{FTT} is the utility token of FTX centralised exchange, mostly used for lowering trading fees on the platform.
Like most other cryptocurrencies, FTT is an inherent Ponzi-type token, meaning that it is not backed by any asset.
The FTT issuers raise money from the first wave of investors with a near-zero cost, just like the initiating debt in any Ponzi game.
FTT may sustain, or may even achieve a rational Ponzi game if it can maintain a relatively stable or increasing market value and be actively traded in a healthy way.
But unfortunately, FTT performs weaker than people previously believed.
FTT does not entitle users to a part of the platform revenue or represent a share in FTX, nor give control over governance decisions or FTX’s treasury.
However, these are not the very causes of its own collapse, the essence of financial distress is loss of confidence.
The change in the mindsets of investors from confidence to pessimism is due to the excessive \textit{leverage} and \textit{divert} operations conducted by FTX and Alameda.

\smallskip
\noindent\textbf{Leverage.} 
Though issued without any reserves, FTT was heavily used as collateral wild west, and thus results in a high leverage to both FTX and Alemeda, which is the second factor facilitating the crash.
Leverage is the ratio of debt to capital or to equity, and we now show how leverage enlarges the available capital in general in Fig.\ref{fig:leverage}.
We denote money markets, including both CeFi platforms and different types of DeFi protocols, are indexed by $\{A, B, ...,N,...\}$.
Starting with the principle of an initial fund, the income of each stage is defined as $\mathcal{I}$, which could be deposit revenue, liquidity mining reward and many so on~\cite{qin2021cefi}.
Thus, the income of $A$ platform is identified as $\mathcal{I}_A$, with a leverage magnifier of $A\%$.
At the same time, we define the life cycle of one type of asset circulating from one platform to another as one round $i$, where $i$ is an integer.
Therefore, we can obtain the combined leverage after $i$ rounds is calculated as,

\begin{equation}
\begin{aligned}
\sum_{i=0}^i \gamma &= 1+A\%+B\%+ ... +N\%.
% \frac{\mathcal{I}_N}{\mathcal{I}_{A}} \\
% &= \frac{\mathcal{I}_N}{\mathcal{I}_{A}}\cdot\frac{\mathcal{I}_A}{\mathcal{I}_{B}}\cdot...\cdot\frac{\mathcal{I}_N}{\mathcal{I}_{N-1}}  \\
% &= 1+x\gamma_A +y\gamma_B + ... + z\gamma_N, \\
% \textit{where} \,\, i & = x+y+z . \\
\end{aligned}
\end{equation}
With such financial approach, FTT was used as collateral to raise a significant amount of money in the market, covering a series of CEXes and DeFi protocols, and thus making its virtual value be magnified thousands of times (\textit{leverage}).
At a high level, FTT plays a very similar role of $\mathsf{M0}$ (currency \& bank reserves) in traditional finance~\cite{van2014bitcoin}, but the actual impacts of FTT and related derivatives make up $\mathsf{M1}$ (money easily used in transactions), or even $\mathsf{M2}$ (money easily used in or converted into use for transactions and real GDP) \cite{mccandless1995some}.

\begin{wrapfigure}{r}{5cm}
	\centering
	\vspace{-10pt}
    \includegraphics[width=0.5\textwidth]{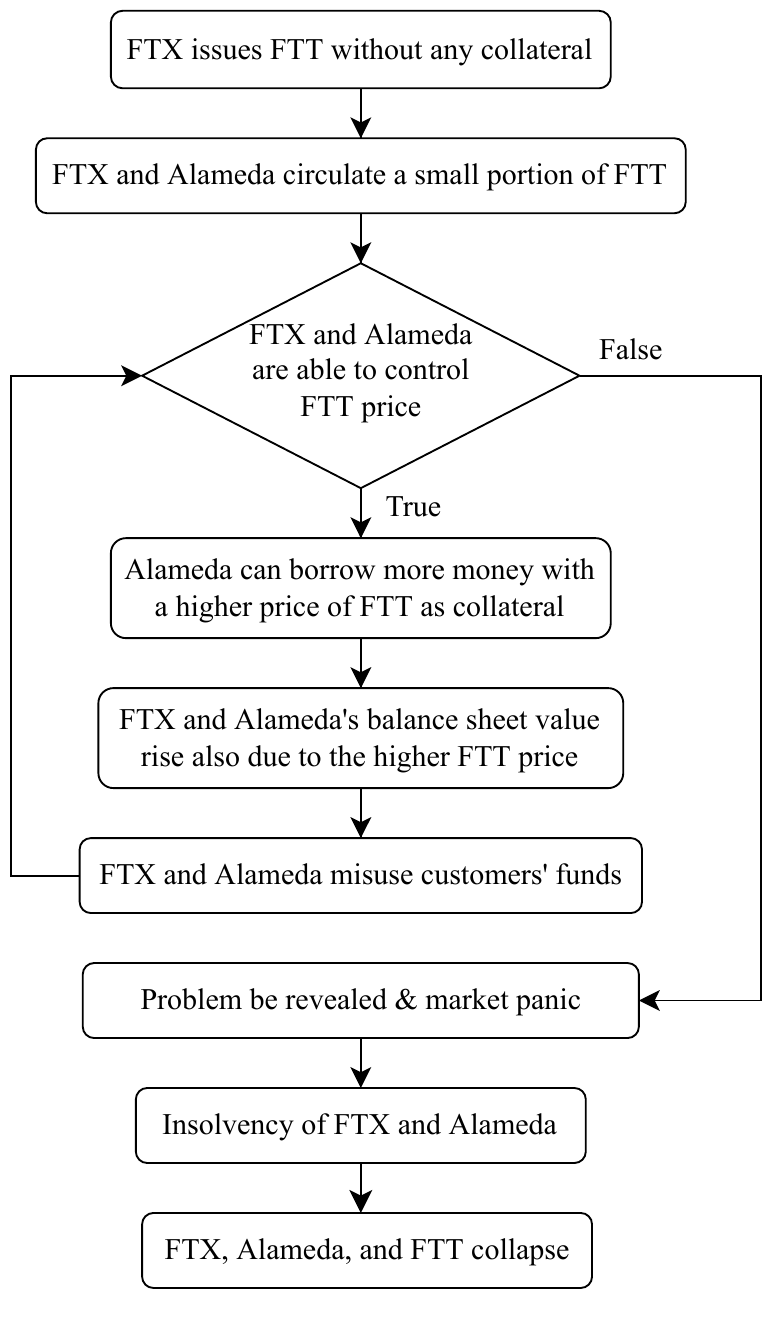}
	\caption{FTX Crash: Logical Flow}
	\label{fig:flow}
	\vspace{-50pt}
\end{wrapfigure}

\smallskip
\noindent\textbf{Diversion.}
In addition to the previous two factors, the misue of customers' funds also facilitates the FTX collapse all along.
As shown in Fig.\ref{fig:flow}, FTX and Alameda only circulate a small portion of the total supply of the FTT, meaning they are able to control the FTT price with funds easily.
On top of this, the vast majority of coins were held by FTX, Alameda, and other associated companies.
They continuously borrow money from investors and entities with FTT as collateral.
To further promote the asset price, FTT managers start to misappropriate the users' reserves (\textit{divert}).
In this sense, the booming of FTX is vigilant.
In retrospect, we know that SBF also spent a lot in donating to political parties, heavy marketing and even bailing out other insolvent companies.
% was just a power move by SBF to detract the public from the insolvency of his own company - FTX.
But there are nearly no external restrictions on how he can use the funds.
% SBF bailing out other insolvent companies, solana, blockfi.
The game ends in a panic when all problems are revealed.
FTX and Alameda were not solvent, and SBF finally paused the withdrawals, marking the official ending of the story.

The FTX crisis was not isolated but a sequel to the previous Luna/UST failure, also the beginning of a new series of collapses.
The previous big collapses of Three Arrows Capital and Terra/Luna meant the market participants were extremely vigilant when it came to any rumors of potential insolvency. The collapse of FTX further damaged the already shaken reputation of the crypto industry.
Besides losing customers' and investors’ funds, the collapse of FTX is already having a knock-on effect on other companies and the overall crypto market.
Solana and BlockFi are among the most severely affected projects.
We refer readers interested in detailed information on this event to the reports by CoinDesk and Nansen~\cite{nansen}\cite{CoinDesk}.
We may also likely expect to see more affected companies being revealed in the near future, and a slow recovery of belief in this market.

%=================================================
\section{Future Directions}
\label{sec:discuss}
%================================================

With lessons learned from the
FTX's crash, we discuss possible ways for better security and sustainability in this area.

\smallskip
\noindent\textbf{Regulation.}
Regulation has been debated since the emergence of cryptocurrencies. On the one side, users moving from traditional finance to crypto space are with strong motivations of pursuing decentralization.
Regulation is naturally placed in the opposite position.
The collapse of FTX raises concerns from users to worry about potential regulatory implications such as heavy regulations or strict policies. However, on the flip side, each time such an influential event happens, a large number of players, unfortunately, run away quickly from crypto.
% Meanwhile, another issue is that the failure of one large CEX will transfer its remaining trafic to another big one, resulting in a smaller range of selections to users.
Meanwhile, the leave of some giants makes it easier for the remaining ones to form a monopoly, making regulatory intervention the only viable solution at this moment.
Therefore, properly introducing regulation into CEXes and even the DeFi world may seem necessary.

\smallskip
\noindent\textbf{Transparency.}
% Transparency is crucial for CEXes. Market participants in centralised markets are fragile when rumors come as they cannot learn in-time news of what has happened.
% Improving transparency can build more confidence for users. 
Transparency is another alternative, which seems more desirable.
Many CEXes already created \textit{proof of reserves} of assets as evidence delivered to the public.
However, the proof of reserve cannot reflect the full picture of exchange solvency such as the liabilities.
CEXes and CeFi-related tools need to rely on external forces to restore markets' confidence.
In contrast, DeFi protocols naturally inherit a series of advantages of on-chain transparency, self-custody\&governance, and fair access for participants \cite{qin2021cefi}.
The logic and rules written in smart contracts are fully transparent that can be publicly checked by anyone. But it still is not an ideal solution due to its high security risks.
Any logic error or loop bugs will be continuously running until the system crashes.
Abilities on self-calibrating\&correcting are absent.
Thus prior security audits are of particular importance to DeFi protocols.
% This can be proved by the fact that during the collapse of FTX, the vast majority of DeFi protocols running on Ethereum and other chains operate correctly as expected, whereas the tragedy still happens. 

%=================================================
\section{Concluding Remarks }
\label{sec:conclusion}
%================================================
% It may take a long for confidence to revive.
 
In this paper, we dig into the root reasons for FTX collapse.
By comparing it with a rational Ponzi model, we identify major violations that break the game rules and as a result, lead to the final crash. Although there are possible ways to avoid similar disasters, none are without their flaws. Further endeavors are required for establishing sustainable token ecosystems. At last, we deliver three pieces of home-taking messages as summaries.

% We observe that the native token \textit{FTT} is an inherent Ponzi-type token, which is issued without any reserves.
% However, FTT was used as collateral to raise a significant amount of money in the market, covering a series of CEXes and DeFi protocols.
% The virtual value has been magnified thousands of times (\textit{leverage}).
% Meanwhile, to further promote the asset price, FTT managers start to misappropriate the users' reserves (\textit{divert}).
% In this sense, the booming of FTX is vigilant and FTX eventually collapsed when potential insolvency comes.

\begin{center}
\tcbset{
        enhanced,
        colback=black!5!white,
        boxrule=0.5pt,
    %    colframe=black!75!black,
        fonttitle=\bfseries
       }
       \begin{tcolorbox}[
       title= Key takeaways,
       lifted shadow={1mm}{-2mm}{3mm}{0.1mm}{black!50!white}
       ]
      \noindent\hangindent 1em\textit{$\triangleright$} Excessive leverage is risky; entities should always place sufficient security deposits while not pursue a dangerous extension of the capital.
      
      \noindent\hangindent 1em\textit{$\triangleright$} The proof-of-reserve standard should be widely adopted in the world of CeFi, in particular CEXes.
      Meanwhile, the transparency of CeFi's corporate balance sheet should be enhanced.
      
      \noindent\hangindent 1em\textit{$\triangleright$} DeFi protocols, powered by blockchain technology, can naturally do better than CeFi in terms of transparency. Its further development, however, still requires rigorous audit and sound governance.

       \end{tcolorbox}
\end{center}

%================================================
{\footnotesize \bibliographystyle{unsrt}
\bibliography{bibb}}

\begin{thebibliography}{10}

\bibitem{cmc}
{CoinMarketCap}.
\newblock {Cryptocurrency Prices, Charts And Market Capitalizations |
  CoinMarketCap}, 2022.
\newblock [Online; accessed 19-Nov-2022].

\bibitem{CoinDesk1}
{CoinDesk}.
\newblock {Divisions in Sam Bankman-Fried’s Crypto Empire Blur on His Trading
  Titan Alameda’s Balance Sheet}.
\newblock {\em
  \url{https://www.coindesk.com/business/2022/11/02/divisions-in-sam-bankman-frieds-crypto-empire-blur-on-his-trading-titan-alamedas-balance-sheet/}},
  2022.

\bibitem{solanaftx}
Khaitan Tarang.
\newblock Solana endures ‘crucible’ as ftx connection deletes 70\% of tv.
\newblock {\em Accessible at
  \url{https://thedefiant.io/solana-test-ftx-crisis}}, 2022.

\bibitem{ftxbacked}
Gilbert Aleksandar.
\newblock Ftx-backed bitcoin and ether tokens depeg on solana.
\newblock {\em Accessible at
  \url{https://thedefiant.io/ftx-files-for-bankruptcy}}, 2022.

\bibitem{usdepartment}
Shumba Camomile.
\newblock Us justice department wants ftx fraud allegations to be investigated.
\newblock {\em Accessible at
  \url{https://www.coindesk.com/policy/2022/12/02/us-justice-department-wants-ftx-fraud-allegations-to-be-investigated/}}.

\bibitem{erc20}
Vogelsteller Fabian and Buterin Vitalik.
\newblock Erc-20 token standard.
\newblock {\em
  \url{https://ethereum.org/en/developers/docs/standards/tokens/erc-20/}}.

\bibitem{wood2014ethereum}
Gavin Wood et~al.
\newblock Ethereum: A secure decentralised generalised transaction ledger.
\newblock {\em Ethereum yellow paper}, 151(2014):1--32, 2014.

\bibitem{basu2014ponzi}
Kaushik Basu.
\newblock The ponzi economy.
\newblock {\em Scientific American}, 310(6):70--75, 2014.

\bibitem{werner2021sok}
Sam~M Werner, Daniel Perez, Lewis Gudgeon, Ariah Klages-Mundt, Dominik Harz,
  and William~J Knottenbelt.
\newblock Sok: Decentralized finance (defi).
\newblock {\em arXiv preprint arXiv:2101.08778}, 2021.

\bibitem{nansen}
Khoo Yong~Li, Leow Sandra, Choe Louisa, Polk Niklas, and Chia Douglas.
\newblock Blockchain analysis: The collapse of alameda and ftx.
\newblock {\em Accessible at
  \url{https://www.nansen.ai/research/blockchain-analysis-the-collapse-of-alameda-and-ftx}},
  2022.

\bibitem{daliaftx}
Dalia Ramirez.
\newblock Ftx crash: Timeline, fallout and what investors should know.
\newblock {\em \url{https://www.nerdwallet.com/article/investing/ftx-crash}},
  2022.

\bibitem{pathforwd}
Jakub.
\newblock Ftx collapse – what is the path forward for crypto.
\newblock {\em \url{https://finematics.com/the-ftx-collapse-explained/}}, 2021.

\bibitem{CoinDesk}
{CoinDesk}.
\newblock {The FTX Downfall: Full Coverage -- Follow the key developments of
  the unraveling of Sam Bankman-Fried’s crypto empire and exchange, FTX.}
\newblock {\em Accessible at
  \url{https://www.coindesk.com/ftx-news-coverage/}}, 2022.

\bibitem{investopedia}
{Nathan, Reiff and Vikki, Velasquez}.
\newblock {The Collapse of FTX: What Went Wrong with the Crypto Exchange?}
\newblock {\em
  \url{https://www.investopedia.com/what-went-wrong-with-ftx-6828447}}, 2022.

\bibitem{forbes}
{Forbes}.
\newblock {The Collapse Of FTX}.
\newblock {\em
  \url{https://www.forbes.com/sites/forbesstaff/article/the-fall-of-ftx/?sh=14bbd8aa7d0c}},
  2022.

\bibitem{wiki}
Wikiwand.
\newblock Bankruptcy of ftx.
\newblock {\em \url{https://www.wikiwand.com/en/Bankruptcy_of_FTX}}, 2022.

\bibitem{jumpto}
Fernau Owen.
\newblock Crypto users jump to defi platforms in wake of ftx’s cefi crash.
\newblock {\em Accessible at \url{https://thedefiant.io/defi-surge-ftx-crash}}.

\bibitem{stablecoins}
Fernau Owen.
\newblock Stablecoins show signs of stabilizing after ftx storm.
\newblock {\em Accessible at
  \url{https://thedefiant.io/stablecoins-show-signs-of-stabilizing-after-ftx-storm}}.

\bibitem{ponzi_gov}
{Investor.gov U.S. SECURITIES AND EXCHANGE COMMISSION}.
\newblock {Ponzi Scheme}, 2022.
\newblock [Online; accessed 19-Nov-2022].

\bibitem{ponzi_rational}
Stephen~A O'Connell and Stephen~P Zeldes.
\newblock Rational ponzi games.
\newblock {\em International Economic Review}, pages 431--450, 1988.

\bibitem{fu2022rational}
Shange Fu, Qin Wang, Jiangshan Yu, and Shiping Chen.
\newblock Rational ponzi games in algorithmic stablecoin.
\newblock {\em arXiv preprint arXiv:2210.11928}, 2022.

\bibitem{blanchard1992dynamic}
Olivier~J Blanchard and Philippe Weil.
\newblock Dynamic efficiency, the riskless rate, and debt ponzi games under
  uncertainty.
\newblock {\em National Bureau of Economic Research Cambridge, Mass., USA},
  1992.

\bibitem{pareto}
Wiki.
\newblock Pareto efficiency.
\newblock {\em Accessible at
  \url{https://www.wikiwand.com/en/Pareto_efficiency}}, 2021.

\bibitem{pitchbook}
{PitchBook}.
\newblock {PitchBook Profile - FTX}, 2022.
\newblock [Online; accessed 19-Nov-2022].

\bibitem{qin2021cefi}
Kaihua Qin, Liyi Zhou, Yaroslav Afonin, Ludovico Lazzaretti, and Arthur
  Gervais.
\newblock Cefi vs. defi--comparing centralized to decentralized finance.
\newblock {\em Crypto Valley Conference on Blockchain Technology (CVCBT)},
  2021.

\bibitem{van2014bitcoin}
Marshall Van~Alstyne.
\newblock Why bitcoin has value.
\newblock {\em Communications of the ACM}, 57(5):30--32, 2014.

\bibitem{mccandless1995some}
George~T McCandless, Warren~E Weber, et~al.
\newblock Some monetary facts.
\newblock {\em Federal Reserve Bank of Minneapolis Quarterly Review},
  19(3):2--11, 1995.

\end{thebibliography}
%================================================

\end{document}